\documentclass{article}
\usepackage{romp}

\usepackage{latexsym}
\usepackage{graphics}
\usepackage{latexsym}
\usepackage{amsmath}
\usepackage{amsfonts}
\usepackage{geometry}
\usepackage{amssymb}

\newtheorem{teo}{Theorem}[section]
\newtheorem{lem}[teo]{Lemma}
\newtheorem{corol}[teo]{Corollary}

\newtheorem{prop}[teo]{Proposition}
\newtheorem{rem}[teo]{Remark}

\newcommand\e{{\rm e}}
\newcommand\p{{\partial}}
\newcommand{\beq}{\begin{equation}}
\newcommand{\eeq}{\end{equation}}

\newcommand{\Z}{{\mathbb{Z}}}
\newcommand{\R}{{\mathbb{R}}}

\newcommand{\C}{{\mathbb{C}}}
\renewcommand{\H}{{\mathcal{H}}}

\renewcommand\det{{\rm det}}
\renewcommand{\Re}{{\rm Re}}
\renewcommand{\Im}{{\rm Im}}

\newcommand{\Sp}{{\rm Sp}}
\newcommand{\Tr}{{\rm Tr}}

\DeclareMathOperator*{\Rz}{Res_0}
\DeclareMathOperator*{\Ru}{Res_1}

\DeclareMathOperator*{\Rk}{Res_k}

\title{Finite temperature quantum field theory on non compact domains and application to delta interactions}

\author{ M. Spreafico \\ Dipartimento di Matematica, Universit\`a di Trento
Povo 38100, Italy. \thanks{On leave from ICMC, Univer\-sidade de S\~{a}o
Paulo, S\~{a}o Carlos, Brasil} \\ {\it mauros@icmc.usp.br} \\
S. Zerbini \\ Dipartimento di Fisica, Universit\`a di Trento\\
e Gruppo Collegato INFN di Trento, Povo 38100, Italy \\ {\it zerbini@science.unitn.it} }

\begin{document}

\maketitle
\begin{abstract} We use relative zeta functions technique of W. Muller \cite{Mul} to extend the classical decomposition of the zeta regularized partition function of a finite temperature quantum field theory on a ultrastatic space-time with
compact spatial section to the case of non compact spatial section. As an application, we study the case of Schr\"{o}dinger operators with delta like potential, as described by Albeverio \& alt. in \cite{AGHH}.
\end{abstract}

\noindent
{2000 {\em Mathematics Subject Classification: 58J52.}

\section{Introduction}

To begin with, we recall that the partition function for a finite
temperature quantum field theory on a ultrastatic space-time with
compact spazial section is constructed as follows. Let $M$ be a
compact Riemannian manifold of dimension $n$, and consider the
product $N=S^1_r\times M$, where $S^1_r$ is the circle of radius
$r=\frac{\beta}{2\pi}$, and $\beta=\frac{1}{T}$ is the inverse of
the temperature. Let $L$ be some non negative self adjoint operator
(typically the Laplacian) acting on some functions space (we shall
deal with scalar fields) defined on $M$ and $H=-\p^2_u+L$. The
canonical partition function at temperature $T$ of this model may be
formally written as $ Z=\det^{-\frac{1}{2}}(\ell^2 H)$, where $\ell$
is some renormalization constant. It is well known that  a rigorous
interpretation to this functional determinat can be given using zeta
function regularization. The zeta function regularization technique
was first introduced by Ray and Singer \cite{RS} to define the
regularized determinant for the Laplacian on  forms, and used by
Hawking \cite{Haw} in order to regularize Gaussian path integrals on
a curved space time, and soon became a fundamental tool in
mathematical physics and may provide a way for regularizing the
partition function of a quantum field theory at finite temperature
on compact domains. Recall that the zeta function of a non negative
self adjoint operator $A$ is defined by (where $\Sp^+ A$ denotes the
positive part of the spectrum of $A$) $\zeta(s;A)=\sum_{\lambda\in
Sp^+ A} \lambda^{-s}$, when $\Re(s)>s_0$ (for some suitable $s_0$),
and by analytic continuation elsewhere. Since $N$ is compact, zero
is not a pole of $\zeta(s;H)$, and using the zeta function, the
regularized functional determinant of $H$ is defined by
\[
\det H=\e^{-\frac{d}{ds}\zeta(s;H)\big|_{s=0}},
\]
and the partition function is
\[
\log Z=\frac{1}{2}\zeta'(0;H)-\frac{1}{2} \zeta(0;H)\log\ell^2.
\]

Introducing the geometric zeta function, namely the zeta function of the restriction $L$ of $H$ to $M$, the following equations hold (we assume here for simplicity that $\ker L=\emptyset$):
\beq\label{ee1}
\zeta(0;H)=
-\beta\Ru_{s=-\frac{1}{2}}\zeta(s;L), \eeq \beq\label{ee2}
\zeta'(0;H)=-\beta\Rz_{s=-\frac{1}{2}} \zeta(s;L)-2\beta(1-\log
2)\Ru_{s=-\frac{1}{2}} \zeta(s;L)-2\log\eta\left(\beta;L\right),
\eeq where the generalized Dedekind eta function for a positive self
adjoint operator $A$ in some Hilbert space $\H(M)$, where $M$ is
compact, is defined by \cite{OS} $ \eta(\tau;A)=\prod_{\lambda\in
\Sp A}\left( 1-\e^{-\tau\sqrt{\lambda}}\right)$, and the notation
$\Rk_{s=s_0} f(s)$ denotes the coefficient of the term
$(s-s_0)^{-k}$ of the Laurent expansion of $f(s)$ at $s=s_0$ (see
for example \cite{BS} pg. 420). This is a classical and well known
result (see for example \cite{gibb77,dowker78}), and we have used
the formulation of \cite{OS} (see also  \cite{ZZ,cognola92} and
\cite{eli1} for an extension), and it leads to the natural question
of a suitable generalization for non compact domains.

In this paper we will try to answer rigorously this question and we
prove a generalization of equations (\ref{ee1}) and (\ref{ee2}) that
holds for a quantum  scalar  field  on a non compact domain. In this
case, we shall consider operators such that the spectrum involved is
not longer only discrete and a continuous contribution appears. With
regard to the treatment of the continuous spectrum, we will follow
the approach of M\"{u}ller \cite{Mul}. However, we should mention
that the introduction of relative traces appeared in the seminal
paper \cite{ulen}, where the so called second virial coefficient,
proportional to the relative trace $\Tr (e^{-\beta
(H_O+V)}-e^{-\beta H_O})$ (here $H_0$ is free Hamiltonian), was
expressed in terms of the trace of scattering matrix, the
Beth-Ulenbech formula. More recently functional determinants in
quantum field theory has been investigated with relative zeta
functions (see for example \cite{dunne}). The mathema\-ti\-cal
counterpart of these approaches in the physical literature makes use
of Krein formula (see \cite{BY}). Beside the natural interest of the
generalization itself, we would like to note that on more general
ground, scattering methods have been applied in the physical
literature (see for example the review \cite{milton07}) in order to
study quantum vacuum effects between material bodies and our result
provides a rigorous justification of these formal approaches.
Furthermore, a motivation is also given by the recently growing
interest in delta interactions, namely a theory described by a
scalar field in a flat space time perturbed by pointlike (uncharged)
"impurities", modeled by delta-like potentials. Since these are
solvable quantum models, it is quite natural to analyze explicitly
these examples, as illustrative application of our results.
Actually, the case of one delta interaction  turns out to be
particularly interesting, since it can be completely solved, and
thus plays the role of the leading example, as the Laplacian on the
circle is the leading example for the compact case.

\section{Relative determinants}
\label{s2}

We introduce in this section the mathematical tools necessary in
order to state our main results. This is essentially based on the
work of M\"{u}ller \cite{Mul}, however, we will reformulate the
approach of M\"{u}ller in terms of the resolvent rather that of the
heat semigroup, because in specific applications we have an explicit
expression for the resolvent function instead than that for the heat
kernel. Anyway, it is well known that one may investigate
equivalently the resolvent of an elliptic operator instead of heat
semigroup.

Let $\H$ be a separable Hilbert space, and let $A$ and $A_0$ be two self adjoint non negative linear operators in $\H$. Suppose that $\Sp A=\Sp_c A\cup \Sp_p A$, where $\Sp_p$ is the point spectrum, and $\Sp_c$ is the continuous spectrum, and we assume both $0$ and $\infty$ are accumulation points of $\Sp A$. It is convenient to split the point spectrum in the null part, $\Sp_p^0 A=\{\lambda_0=0\}$, that has finite multiplicity, plus the positive part, $\Sp_p^+A=\{\lambda_j\}_{j=1}^J$, where each eigenvalue is counted according to multiplicity. Let $\H=\H_c\oplus\H_p$ be the orthogonal decomposition into the subspaces that correspond to the continuous and the point spectrum of $A$, respectively, and let $A_c$ and $A_p$ denote the restrictions of $A$ to $\H_c$ and $\H_p$, respectively.
Let $R(\lambda,T)=(\lambda I-T)^{-1}$ denotes the resolvent of the operator $T$, and $\rho(T)$ the resolvent set.
Then, we introduce the following two sets of conditions. First, we assume that the sequence $\Sp_p^+ A$ is a totally regular sequence of spectral type with finite exponent $s_0$, as defined in \cite{Spr9}. This implies that the following conditions hold:
\begin{enumerate}
\item[(A.1)] The operator $R(\lambda,A_p)$ is of trace class for all $\lambda\in \rho(A_p)$;
\item[(A.2)] as $\lambda\to \infty$ in $\rho(A_p)$, there exists an asymptotic expansion of the form:
\[
\Tr R(\lambda,A_p)-\dim\ker A_p\frac{1}{\lambda}\sim \sum_{j=0}^\infty \sum_{k=0}^{K_j} a'_{j,k} (-\lambda)^{\alpha'_j} \log ^k (-\lambda),
\]
where $-\infty<\dots<\alpha'_1<\alpha'_0\leq s_0-1$, and $\alpha'_j\to -\infty$, for large $j$, and $a'_{j,k}=0$ for $k>0$;
\item[(A.3)] as $\lambda\to 0$, there exists an asymptotic expansion of the form
\[
\Tr R(\lambda,A_p)-\dim\ker A_p\frac{1}{\lambda}\sim \sum_{j=0}^\infty b'_{j} (-\lambda)^{\beta'_j},
\]
where $0=\beta'_0<\beta'_1<\dots$, and $\beta'_j\to +\infty$, for large $j$.
\end{enumerate}
Second, we assume the following conditions on the pair $(A_c,A_0)$:
\begin{enumerate}
\item[(B.1)] the operator $R(\lambda,A_c)-R(\lambda,A_0)$ is of trace class for all $\lambda\in \rho(A_c)\cap\rho(A_0)$;
\item[(B.2)] as $\lambda\to \infty$ in $\rho(A_c)\cap\rho(A_0)$, there exists an asymptotic expansion of the form:
\[
\Tr (R(\lambda,A_c)-R(\lambda,A_0))\sim \sum_{j=0}^\infty \sum_{k=0}^{K_j} a_{j,k} (-\lambda)^{\alpha_j} \log ^k (-\lambda),
\]
where $-\infty<\dots<\alpha_1<\alpha_0$, $\alpha_j\to -\infty$, for large $j$, and $a_{j,k}=0$ for $k>0$;
\item[(B.3)] as $\lambda\to 0$, there exists an asymptotic expansion of the form
\[
\Tr (R(\lambda,A_c)-R(\lambda,A_0))\sim \sum_{j=0}^\infty b_{j} (-\lambda)^{\beta_j},
\]
where $-1\leq\beta_0<\beta_1<\dots$, and $\beta_j\to +\infty$, for large $j$.
\end{enumerate}

We introduce the further consistency condition (that will be always tacitely assumed)
\begin{enumerate}
\item[(C)] $\alpha_0<\beta_0$.
\end{enumerate}

By results of \cite{Spr9}, it follows that the zeta function of the
operator $A_p$ is well defined by the uniformly convergent series
\[
\zeta(s;A_p)=\sum_{j=1}^\infty \lambda_j^{-s},
\]
when $\Re(s)>s_0\geq \alpha'_0+1$, and  by analytic continuation elsewhere.
In particular, the heat semigroup $\e^{- tA_p}$ is of trace class, and the following equations hold:
\beq\label{p1}
\Tr \e^{-A_p t}-\dim\ker A_p =\frac{1}{2\pi i}\int_{\Lambda_{\theta,-a}} \e^{-\lambda t} \left(\Tr R(\lambda,A_p)-\dim\ker A_p\frac{1}{\lambda}\right) d\lambda,
\eeq
where the Hankel type contour is $\Lambda_{\theta,-a}=\left\{\lambda\in\C~|~|\arg(\lambda+a)=\frac{\theta}{2}\right\}$, oriented counter clockwise, with some fixed $a>0$, $0<\theta<\pi$.
\beq\label{p2}
\zeta(s;A_p)=\frac{1}{\Gamma(s)}\int_0^\infty t^{s-1}\left( \Tr\e^{-tA_p}-\dim\ker A_p\right)dt.
\eeq

We can prove similar results for the relative heat semigroup and the relative zeta function, using \cite{Mul}. For,  we introduce the following lemma.

\begin{lem} \label{l1} If the pair of non negative self adjoint operators $(T,T_0)$ satisfies conditions (B.1)-(B.3), then it satisfies the conditions (1.1)-(1.3) of \cite{Mul}.
\end{lem}
\noindent{\underline{Proof}} The proof that conditions (B.1) and (B.2) imply conditions (1.1) and (1.2) of \cite{Mul} follows from the following equation
\beq\label{el1}
\e^{-T t}-\e^{-T_0 t}=\frac{1}{2\pi i}\int_{\Lambda_{\theta,-a}} \e^{-\lambda t} \left(R(\lambda,T)-R(\lambda,T_0)\right) d\lambda.
\eeq
and 2.2 of \cite{Spr9}, respectively. Next, assume (B.3). Then, for any fixed $\beta_J$,
\[
\left|\Tr (R(\lambda,T)-R(\lambda,T_0))-b_j (-\lambda)^{\beta_{J-1}}\right|\leq K |-\lambda|^{\beta_J}.
\]

We can use this bound for the remainder in order to obtain (1.3) of \cite{Mul}. For, using the expansion given by condition (B.3) of the difference of the resolvents in equation (\ref{el1}), the remainder is
\[
r_{J}(t)=\frac{1}{2\pi i}\int_{\Lambda_{\theta,-a}} \e^{-\lambda t}\left(\Tr (R(\lambda,T)-R(\lambda,T_0))-\sum_{j=0}^J b_j (-\lambda)^{-\beta_j}\right) d\lambda,
\]
and thus it satisfies the bound
\[
|r_J(t)|\leq K \int_{|\Lambda_{\theta,-a}|}|\e^{-\lambda}||(-\lambda)^{\beta_J}| |d\lambda| t^{-\beta_J-1},
\]
where the integral is a finite constant. \rule{5pt}{5pt}

Therefore, assuming conditions (B.1)-(B.3) for the pair of nonnegative self adjoint operators $(A_c,A_0)$, all the results of \cite{Mul} hold, and in particular we can define the relative zeta function for the pair $(A_c,A_0)$ by the following equation
\beq\label{c2}
\zeta(s;A_c,A_0)=\frac{1}{\Gamma(s)}\int_0^\infty t^{s-1} \Tr\left(\e^{-tA_c}-\e^{-tA_0 }\right)dt,
\eeq
when $\alpha_0+1<\Re(s)<\beta_0+1$, and by analytic continuation elsewhere. Back to the pair $(A,A_0)$, note that
\[
\Tr \left(\e^{-tA}-\e^{-tA_0}\right)=\Tr \e^{-tA_p }+\Tr \left(\e^{-tA_c }-\e^{-tA_0}\right),
\]
and the problem decomposes additively into the two terms arising from the pure point and the continuous spectrum, namely $\zeta(s;A,A_0)=\zeta(s;A_p)+\zeta(s;A_c,A_0)$.
Thus, if we define the regularized relative determinant of the pair of operators $(A,A_0)$ by
\[
\det (A,A_0)=\e^{-\frac{d}{ds}\zeta(s;A,A_0)\big|_{s=0}},
\]
then, we have the decomposition
\[
\det(A,A_0)=\det(A_p)\det(A_c,A_0),
\]
and the two regularizations can be treated independently.
This suggests to introduce the following definition for the zeta
regularized partition function of a model described by the operator
$A$, under the assumption that there exists a second operator $A_0$
such that (using the above decomposition) the operator $A_p$
satisfies assumptions (A.1)-(A.2) and the pair of operators
$(A_c,A_0)$ satisfies assumptions (B.1)-(B.3):
\beq\label{partition2} \log Z =\frac{1}{2}\zeta'(0;A_p)-\frac{1}{2}
\zeta(0;A_p)\log \ell^2+\frac{1}{2}\zeta'(0;A_c,A_0)-\frac{1}{2}
\zeta(0;A_c,A_0)\log \ell^2, \eeq

This is the natural generalization of the classical zeta regularization technique to the relative case.
We conclude this section with a technical result.

\begin{lem}\label{ll} Assume the non negative self adjoint operator $T$ decomposes additively as sum of two non negative self adjoint commuting operators $T_1$ and $T_2$, where $\e^{-tT_1}$ is compact trace class and satisfies an expansion for small $t$ as $\sum_{j=0}^\infty \sum_{k=0}^{K_j} c_{j,k} t^{\gamma_j} \log ^k t$, with $-\infty<\gamma_0<\gamma_1<\dots$,  $\gamma_j\to +\infty$, for large $j$, and $c_{j,k}=0$ for $k>0$.
Then, if there exists an operator $T_0$ such that conditions (B.1)-(B.3) hold for the pair $(T_2,T_0)$,  then conditions (1.1)-(1.3) of \cite{Mul} hold for the pair $(T,T_1+T_0)$, and viceversa. In particular, the following equation holds
\[
{\rm Tr}\left(\e^{-tT}-\e^{-t(T_1+T_0)}\right)={\rm Tr}\e^{-tT_1}{\rm Tr}\left(\e^{-tT_2 }-\e^{-tT_0}\right).
\]
\end{lem}
\noindent\underline{Proof} By standard properties of the heat semigroup
\[
\e^{-tT}-\e^{-t(T_1+T_0)}=\e^{-tT_1}\left(\e^{-tT_2 }-\e^{-tT_0}\right).
\]
Suppose (B.1) hold for $(T_2,T_0)$. Then, $\e^{-tT_2 }-\e^{-tT_0}$ is of trace class by Lemma \ref{l1}, and the above equation implies that $\e^{-tT }-\e^{-t(T_1+T_0)}$ is of trace class. Therefore, the equation given in the statement of the lemma  holds. This implies that, if (B.2) and (B.3) hold for $(T_2,T_0)$ and (A.2) holds for $T_1$, then (1.2)-(1.3) of \cite{Mul} hold for $(T,T_1+T_0)$.
The proof of the converse is similar.
\rule{5pt}{5pt}

\section{Relative partition function}
\label{s22}

Let $M$ be a smooth Riemannian manifold of dimension $n$, and
consider the product $N=S^1_\frac{\beta}{2\pi}\times M$, where
$S^1_r$ is the circle of radius $r$. Let $\xi$ be a complex line
bundle over $N$, and $L$ a self adjoint non negative linear operator
on the Hilbert space $\H(M)$ of the $L^2$ sections of the
restriction of $\xi$ onto $M$, with respect to some fixed metric $g$
on $M$. Let $H$ be the self adjoint non negative operator
$H=-\p^2_u+L$, on the Hilbert space $\H(N)$ of the $L^2$ sections of
$\xi$, with respect to the product metric  $du^2\oplus g$  on $N$,
and with periodic boundary conditions on the circle. Assume that
there exists a second operator $L_0$ defined on $\H(M)$, such that
the pair $(L,L_0)$ satisfies the assumptions (B.1)-(B.3) of Section
\ref{s2} (since we have seen that the problem decomposes additively
in point and continuous part, we assume here without loss of
generality that the point spectrum is empty). Then, by Lemma
\ref{ll}, it follows that there exists a second operator $H_0$
defined in $\H(N)$, such that the pair $(H,H_0)$ satisfies those
assumptions too. Under these requirements, we introduce the relative
zeta regularized partition function of the model described by the
pair of operators $(H,H_0)$ using equation (\ref{partition2}), and
we can prove the following result.

\begin{prop}\label{pr1} Let $L$ be a non negative self adjoint operator on $M$, and $H=-\p^2_u+L$, on $S^1_r\times M$ as defined above. Assume there exists an operator $L_0$ such that the pair $(L,L_0)$ satisfies conditions (B.1)-(B.3). Then,
\begin{align*}
\zeta(0;H,H_0)&=
-\beta\Ru_{s=-\frac{1}{2}}\zeta(s;L,L_0),\\
\zeta'(0;H,H_0)&=-\beta\Rz_{s=-\frac{1}{2}}
\zeta(s;L;L_0)-2\beta(1-\log 2)\Ru_{s=-\frac{1}{2}}
\zeta(s;L,L_0)-2\log\eta(\beta ;L,L_0),
\end{align*}
where $H_0=-\p^2_u+L_0$ and the relative Dedeckind eta function is
defined by
\begin{align*}
\log\eta(\tau;L,L_0)&=\int_0^\infty \log\big(1-\e^{-\tau v}\big) e(v;L,L_0) dv,\\
e(v;L,L_0)&=\frac{v}{\pi i}\lim_{\epsilon\to 0^+}\left(r(v^2\e^{2i\pi-i\epsilon};L,L_0)-r(v^2\e^{i\epsilon};L,L_0)\right),\\
r(\lambda;L,L_0)&=\Tr(R(\lambda,L)-R(\lambda,L_0)).
\end{align*}
\end{prop}
\noindent\underline{Proof} Since $(L,L_0)$ satisfies (B.1)-(B.3), by Lemma \ref{ll} $(H,H_0)$ satisfies (1.1)-(1.3) of \cite{Mul} and the zeta function is defined by
\[
\zeta(s;H,H_0)=\frac{1}{\Gamma(s)}\int_0^\infty t^{s-1} {\rm Tr}\big(\e^{-tH}-\e^{-tH_0 }\big)dt,
\]
when $\alpha_0+1<\Re(s)<\beta_0+1$. By Lemma \ref{ll}
\begin{align*}
{\rm Tr}\left(\e^{-Ht}-\e^{-H_0t}\right)=&\sum_{n\in \Z}\e^{-\frac{n^2}{r^2}t} {\rm Tr}\left(\e^{-tL }-\e^{-tL_0 }\right),
\end{align*}
and hence, using the well known Jacobi summation formula
we obtain
\beq\label{zeta}
\begin{aligned}
\zeta(s;H,H_0)=&\frac{1}{\Gamma(s)}\int_0^\infty t^{s-1} \sum_{n\in \Z}\e^{-\frac{n^2}{r^2} t}{\rm Tr}\left(\e^{-tL}-\e^{-tL_0 }\right)dt\\
=&\frac{\sqrt{\pi}r}{\Gamma(s)}\int_0^\infty t^{s-\frac{1}{2}-1} {\rm Tr}\left(\e^{-tL}-\e^{-tL_0 }\right)dt\\
&+\frac{2\sqrt{\pi}r}{\Gamma(s)}\int_0^\infty t^{s-\frac{1}{2}-1} \sum_{n=1}^\infty\e^{-\frac{\pi^2 r^2 n^2}{t}}{\rm Tr}\left(\e^{-tL}-\e^{-tL_0 }\right)dt\\
=&\frac{\sqrt{\pi}r}{\Gamma(s)}\Gamma\left(s-\frac{1}{2}\right) \zeta\left(s-\frac{1}{2};L,L_0\right)\\
&+\frac{2\sqrt{\pi}r}{\Gamma(s)}\sum_{n=1}^\infty\int_0^\infty t^{s-\frac{1}{2}-1} \e^{-\frac{\pi^2 r^2n^2}{t}}{\rm Tr}\left(\e^{-tL}-\e^{-tL_0 }\right)dt\\
=&z_1(s)+z_2(s).
\end{aligned}
\eeq

The first term, $z_1(s)$, can be expanded near $s=0$, and this gives the result stated. In fact, by Proposition 1.1 of \cite{Mul}, $\zeta(s;H,H_0)$ is regular at $s=0$, and this implies that the pole of $\zeta(s;L,L_0)$ at $s=-\frac{1}{2}$ is simple. To deal with the second term, since $(L,L_0)$ satisfies (B.1)-(B.3), we can write
\[
\Tr\left(\e^{-tL }-\e^{-tL_0 }\right)=\frac{1}{2\pi i}\int_{\Lambda_{\theta,-a}} \e^{-\lambda t} \Tr(R(\lambda,L)-R(\lambda,L_0))d\lambda.
\]

Now, it is convenient to change the spectral variable to $k=\lambda^{\frac{1}{2}}$, with the principal value of the square root, i.e. with $0<\arg k<\pi$. Then,
\[
{\rm Tr}\left(\e^{-tL}-\e^{-tL_0  }\right)=\frac{1}{\pi i}\int_\gamma \e^{-k^2 t} \Tr(R(k^2,L)-R(k^2,L_0))k d k,
\]
where $\gamma$ is the line $k=-ic$, for some $c>0$. Writing
$k=v\e^{i\theta}$, and $r(\lambda;L,L_0)=\Tr(R(\lambda,L)-R(\lambda,L_0))$, a standard computation leads to
\begin{align}
\label{eea}{\rm Tr}\left(\e^{-tL }-\e^{-tL_0 }\right)&=\int_0^\infty \e^{-v^2 t}e(v;L,L_0) dv,\\
\label{eeb}\zeta(s;L,L_0) &=\int_0^\infty v^{-2s}e(v;L,L_0) dv .
\end{align}
where we have introduced  the trace of the relative spectral
measure  \beq\label{spe} e(v;L,L_0)= \lim_{\epsilon\to 0^+}
\frac{v}{\pi i}(r((v^2\e^{2i\pi-i\epsilon}
;L,L_0)-r(v^2\e^{i\epsilon};L,L_0)), \eeq associated to the pair of
operators $(L,L_0)$.

As a result, the second term, $z_2(s)$, of equation (\ref{zeta}) becomes
\begin{align*}
z_2(s)=&\frac{2\sqrt{\pi}r}{\Gamma(s)}\sum_{n=1}^\infty\int_0^\infty t^{s-\frac{1}{2}-1} \e^{-\frac{\pi^2n^2r^2}{t}}\int_0^\infty \e^{-v^2 t} e(v;L,L_0) d vdt,\\
\end{align*}
and we can do the $t$ integral using for example \cite{GZ} 3.471.9. We obtain
\beq\label{po}\begin{aligned}
z_2(s)=&\frac{4\sqrt{\pi} r}{\Gamma(s)}\sum_{n=1}^\infty\int_0^\infty \left(\frac{\pi n r}{v}\right)^{s-\frac{1}{2}}K_{s-\frac{1}{2}}(2\pi n rv) d v.\\
\end{aligned}
\eeq

Since the Bessel function is analytic in its parameter, regular at $-\frac{1}{2}$, and $K_{-\frac{1}{2}}(z)=\sqrt{\frac{\pi}{2z}}\e^{-z}$, equation (\ref{po}) gives the formula for the analytic extension of the zeta function $\zeta(s;H,H_0)$ near $s=0$. We obtain
\begin{equation}
z_2(0)=0\,, \quad
z_2'(0)=-2\int_0^\infty \log\left(1-\e^{-2\pi rv}\right)e(v;L,L_0) d v,
\end{equation}
and the integral converges by assumptions (B.2) and (B.3), and equation (\ref{spe}) for the trace of the spectral measure.
This completes the proof.

\rule{5pt}{5pt}

Note that equation (\ref{partition2}) is the natural generalization of the zeta function technique from the absolute (compact) to the relative (non compact) case, and that Proposition \ref{pr1} extends the main result of the absolute case (see equations (\ref{ee1}) and (\ref{ee2})) to the relative one. Namely, the partition function of the model described above satisfies the equation stated in Corollary \ref{cc1}, that follows from equation (\ref{partition2}) and Proposition \ref{pr1}.

\begin{corol} \label{cc1}
\[
\log Z=\beta\left(\log
2\ell-1\right)\Ru_{s=-\frac{1}{2}}\zeta(s;L,L_0)-\frac{\beta}{2}\Rz_{s=-\frac{1}{2}}\zeta(s;L,L_0)-\log\eta\left(\beta;L,L_0\right).
\]
\end{corol}

Next, we show that a further feature of the compact case, namely the behavior for small temperature, extends to the non compact case.

\begin{corol} \label{cc2} For large $\beta$
\[
\log Z=-E_{vacuum}\beta+O\left(\beta^{-\epsilon}\right),
\]
with some $\epsilon>0$, and where $E_{vacuum}$ is the vacuum energy
\[
E_{vacuum}=-\left(\log
2\ell-1\right)\Ru_{s=-\frac{1}{2}}\zeta(s;L,L_0)+\frac{1}{2}\Rz_{s=-\frac{1}{2}}\zeta(s;L,L_0).
\]
\end{corol}
\noindent\underline{Proof} By definition
\[
E_{vacuum}=-\lim_{\beta\to +\infty} \partial_\beta \log Z.
\]

The result follows from the equation given in Corollary
\ref{cc1}, once we show that $\log\eta\left(\beta;L,L_0\right)=O(\beta^{-\epsilon})$, for large $\beta$. For, recall the definition of the Dedekind eta function
\[
\log\eta(\tau;L,L_0)=\int_0^\infty \log\big(1-\e^{-\tau v}\big)
e(v;L,L_0) dv.
\]

We split the integral at $v=1$. Since the above integral converges
uniformly for large $v$, the $\int_1^\infty$ is  $O(\e^{-\tau})$ for
large $\tau$. For the other integral, we use the expansion of the
trace of the relative resolvent $r(\lambda;L,L_0)$ for small
$\lambda$ assumed by condition (B.3), to obtain from the definition
of the relative spectral measure, equation (\ref{spe}), that
$e(v;L,L_0)=O(v^{1+2\beta_j})$, where $\beta_j$ is the first non
integer exponent in (B.3).
\rule{5pt}{5pt}


\begin{rem} Equation (\ref{eeb}) defines the relative zeta function when $\alpha_0+1<\Re(s)<\beta_0+1$.
In general, the meromorphic continuation of this quantity has a simple pole at $s=-1/2$ and the formal definition of the vacuum energy
\[
E_{vacuum}=\int_0^\infty v e(v;L,L_0) dv,
\]
is meaningless. Our formula  in Corollary \ref{cc2} provides a rigorous
regularization scheme for this formal definition.
\end{rem}

\section{Zeta regularized partition function for delta interactions}
\label{s3}

We analyze in this section two natural applications of the method
presented in Section \ref{s2}. The geometry of our model is given by
a scalar field in the three dimensional flat space interacting with
one or two external fields described by delta like potentials, thus
the geometric operator describing our model is formally
$L=-\Delta-\mu_0 \delta(0)-\mu_1\delta(a)$, where $\Delta$ is the
Laplace operator in $\R^3$, $\mu_j$ are real constants (the strength
of the interactions), and $a$ a fixed point in $\R^3$. Models of
this type have been studied by different authors (see \cite{bordag}
\cite{irina}). In particular, in the case of a one point
interaction, a rigorous definition has been also obtained by Green's
function approach, and formulas for the heat kernel has been given
\cite{ST} \cite{Par} \cite{Sol}. However, a unified approach valid
for finitely many points interaction, was presented   by Albeverio
et al. in  \cite{AGHH}, using Fourier transform, a method first used
in \cite{fad}. We will use this approach.

\subsection{One point interaction in three dimensions}
\label{s3.1} The concrete geometric operator describing our model is $L=-\Delta_{\alpha}$, where $-\Delta_{\alpha}$ is defined in Theorem I.1.1.2 of \cite{AGHH} by the resolvent with the following kernel
\beq\label{kk}
{\rm ker}(x,x',(\lambda I+\Delta_{\alpha})^{-1})=-G_k(x-x')-\frac{1}{\alpha-\frac{i k}{4\pi}}G_k^2(x),
\eeq
with $\lambda=k^2\in \rho(-\Delta_{\alpha,a})$, $\Im k>0$,  $\alpha$ is a real parameter related to the strength $\mu_0$ (we have $\mu_1=0$ in the present case) \cite{AGHH} II.1.1.30, and the free Green function is:
\begin{align*}
G_k(x)&=\frac{\e^{ik|x|}}{4\pi |x|}.
\end{align*}

Note that the case $\alpha=\infty$ corresponds to the negative free Laplace operator $-\Delta=-\Delta_{\infty}$.
By \cite{AGHH} Theorem I.1.1.4 the spectrum of $-\Delta_{\alpha}$ is purely absolutely continuous ${\rm Sp}(-\Delta_{\alpha})=[0,\infty)$, if $\alpha\geq 0$, while has one negative eigenvalue, $\lambda=-(4\pi \alpha)^2$, if $\alpha< 0$.

The complete operator describing our model is
$H=-\partial^2_u-\Delta_{\alpha}$, and, because of  the above result
on the spectrum of $-\Delta_\alpha$, we assume  $\alpha\geq 0$.
Proceeding as in Section \ref{s22}, we introduce the unperturbed
operator $H_0=-\partial_u^2-\Delta$, and we consider the pair of
operators $(H,H_0)$. The partition function of our model is given by
equation (\ref{partition2}), without the part arising from the point
spectrum, and we need to study the analytic continuation of  the
relative zeta function $\zeta(s;H,H_0)$. We first check that the
requirements (B.1)-(B.3) of Section \ref{s2} are satisfied by the
pair of operators $(L,L_0)=(-\Delta_\alpha, -\Delta)$, and this is
true as a particular instance of a general class of pairs of
operators considered in Section 4.1 of \cite{Mul} (and references
therein for this particular type of potential) or in Section 1.6 of
\cite{BY}. However, note that we will be able to verify directly
conditions (B.1)-(B.3). Next, using equation (\ref{kk}), the
difference of the kernels of the resolvents of the geometric
operators is
\begin{align*}
{\rm ker}(x,x',R(\lambda,-\Delta_{\alpha}))-{\rm ker}(x,x',R(\lambda, -\Delta))
&=-\frac{\e^{2ik|x|}}{4\pi |x|^2 (4\pi \alpha-i k)},
\end{align*}
and is class trace, since it follows
\begin{align*}
{\rm Tr}(R(\lambda,-\Delta_{\alpha})-R(\lambda,-\Delta))=&\frac{1}{2ik (4\pi \alpha-i k)}.
\end{align*}

A further simple computation gives the trace of the relative spectral measure
\beq\label{spectral}
\begin{aligned}
e(v;-\Delta_\alpha,-\Delta)
&=\frac{4\alpha}{(4\pi\alpha)^2+v^2},
\end{aligned}
\eeq
and the following  formulas for the main geometric spectral functions:
\begin{align}
\label{l1.1} {\rm Tr}\big(\e^{-t(-\Delta_\alpha)}-\e^{-t(-\Delta)}\big)=&\frac{\e^{(4\pi\alpha)^2t}}{2}\left(1-\Phi(4\pi\alpha\sqrt{t})\right),\\
\label{l1.2} \zeta(s;-\Delta_\alpha,-\Delta)
=&\frac{1}{2}\frac{(4\pi\alpha)^{-2s}}{\cos\pi s},\\
\label{l1.4} \eta(\tau;-\Delta_\alpha,-\Delta)=&\log\Gamma(2\alpha\tau)+\frac{1}{2}\log2\alpha\tau-2\alpha\tau(\log2\alpha\tau-1)-\frac{1}{2}\log2\pi.
\end{align}

The formula in equation (\ref{l1.1}) follows from the definition, since using equations (\ref{eea}) and (\ref{spectral}),
\begin{align*}
{\rm Tr}\big(\e^{-(-\Delta_\alpha)t}-\e^{-(-\Delta)t}\big)
&=4\alpha\int_0^\infty  \frac{\e^{-v^2 t}}{(4\pi\alpha)^2+v^2}dv,\\
\end{align*}
and next we can apply \cite{GZ} 3.363.2 (the probability integral
function is defined accordingly to \cite{GZ} 8.250 -  recall
$\alpha$ is non negative). Note that the  integral representation
for the trace of the difference of the heat operators given in
equation (\ref{l1.1}) verifies the conditions (1.1)-(1.3) of
\cite{Mul} for the pair of operators $(-\Delta_\alpha, -\Delta)$.
The formula in equation (\ref{l1.2}) follows using equations
(\ref{eeb}), and (\ref{spectral}). The same result also follows
using the formula in equation (\ref{eea}), and the previous result
for the trace of the difference of the heat semigroups, under the
condition that $\Re(s)>0$, and using \cite{GZ} 6.286.1. The formula
in equation (\ref{l1.4}) follows by definition and \cite{GZ}
4.319.1.

Expanding the formula in equation (\ref{l1.2}) near $s=0$, we obtain
\begin{align*}
\Ru_{s=-\frac{1}{2}}\zeta(s;-\Delta_\alpha,-\Delta)=2\alpha,\hspace{30pt}
\Rz_{s=-\frac{1}{2}}\zeta(s;-\Delta_\alpha,-\Delta)=-4\alpha\log 4\pi\alpha.
\end{align*}

Using these results  in Corollary \ref{cc1}, we obtain the explicit
formula for the partition function

\begin{align*}
\log Z=&2\left(\log4\pi\alpha\ell-1\right)\alpha\beta-\log\eta\left(\beta;-\Delta_\alpha,-\Delta\right)\\
=&2\left(\log4\pi\alpha\ell-1\right)\alpha\beta-\log\Gamma\left(2\alpha\beta\right)-\frac{1}{2}\log2\alpha\beta
+2\alpha\beta(\log2\alpha\beta-1)+\frac{1}{2}\log2\pi.
\end{align*}

Note that, using the classical expansion for the Gamma function,
this result is consistent with Corollary \ref{cc2}. The regularized
vacuum energy follows immediately from the above expression.

\subsection{Two point interactions in three dimensions}
\label{s3.2} The concrete geometric operator describing our model is $L=-\Delta_{\alpha,a}$, where $-\Delta_{\alpha,a}$ is defined in Theorem II.1.1.1 of \cite{AGHH}, by the resolvent with the following integral kernel
\[
{\rm ker}(x,x',(\lambda I+\Delta_{\alpha,a})^{-1})=-G_k(x-x')-\sum_{j,l=0}^1 \Gamma_{\alpha,a}^{-1}(k)_{j,l}G_k(x-a_j) G_k(x'-a_l),
\]
with $\lambda=k^2\in \rho(-\Delta_{\alpha,a})$, $\Im k>0$,  and where the $\alpha_j$ are real parameters (see \cite{AGHH} II.(1.1.25)), and
\begin{align*}
\Gamma_{\alpha,a}(k)&=\left(\begin{array}{cc}\alpha_0-\frac{ik}{4\pi}&-G_k(a)\\
-G_k(a)&\alpha_1-\frac{ik}{4\pi}\end{array}\right).
\end{align*}

Note that the case $\alpha_j=\infty$ corresponds to the negative free Laplace operator $-\Delta=-\Delta_{\infty,a}$, and $\alpha_1=\infty$ to the case considered in Section \ref{s3.1}.
By \cite{AGHH} Theorem I.1.1.4 the spectrum of $-\Delta_{\alpha,a}$ is purely absolutely continuous $Sp(-\Delta_{\alpha,a})=[0,\infty)$, plus at most two negative eigenvalues. The eigenvalues are present if $\det \Gamma_{\alpha,a}(k)=0 $
for $\Im k >0$. An explicit analysis (see also the end of Section II.1.1 of \cite{AGHH}) shows that the condition necessary in order to have a purely continuous spectrum is $4\pi^2\alpha_0\alpha_1 a^2\geq 1$. We will proceed assuming this condition.

The unperturbed geometric operator is $-\Delta$, and the fact that
the pair $(-\Delta_{\alpha,a},-\Delta)$ satisfies conditions
(B.1)-(B.3) follows as in Section \ref{s3.1}. The difference of the
resolvents has trace
\begin{align*}
{\rm Tr} (R(k^2,-\Delta_{\alpha,a})-R(k^2,-\Delta))
=&\frac{a^2}{ika}\frac{2\pi(\alpha_0+\alpha_1)a-ika+\e^{2ika}}{\left(4\pi\alpha_0a-ika\right)\left(4\pi \alpha_1a-ika\right)-\e^{2ika}}.
\end{align*}

This allows to write a formula for the trace of the relative spectral measure. Using the definition in equation (\ref{spe}), we obtain
\begin{align*}
e(v;-\Delta_{\alpha,a},&-\Delta)\\
&=\frac{a}{\pi}\left(\frac{2\pi(\alpha_0+\alpha_1)a-iav+\e^{2iav}}{a^2\left(4\pi\alpha_0-iv\right)\left(4\pi
\alpha_1-iv\right)-\e^{2iav}}+\frac{2\pi(\alpha_0+\alpha_1)a+iav+\e^{-2iav}}{a^2\left(4\pi\alpha_0a+iv\right)\left(4\pi
\alpha_1a+iv\right)-\e^{-2iav}}\right).
\end{align*}

Note that in the limit case $\alpha_1\to \infty$ the relative
spectral measure $e(v;-\Delta_{\alpha,a},-\Delta)$ reduces smoothly
to the one $e(v;-\Delta_{\alpha_0},-\Delta)$, considered in Section
\ref{s3.1}.

The  formula for the trace of the relative spectral measure allows to compute all the quantities appearing in the Proposition \ref{pr1}, and therefore to obtain an explicit result for the partition function using Corollary \ref{cc1}.
For, note that the function $e(v;-\Delta_{\alpha,a},-\Delta)$ is a
smooth function, as it is the quotient of powers and trigonometric
functions.
To compute the values of the residue and of the finite part of the
zeta function $\zeta(s;-\Delta_{\alpha,a},-\Delta)$ at
$s=-\frac{1}{2}$, we use the expansions of
$e(v;-\Delta_{\alpha,a},-\Delta)$ for small and large $v$. For small
$v$,
\[
e(v;-\Delta_{\alpha,a},-\Delta)=\frac{a}{\pi}\frac{4\pi(\alpha_0+\alpha_1)a+2}{16\pi^2\alpha_0\alpha_1
a^2-1}+O(v),
\]
while for large $v$
\[
e(v;-\Delta_{\alpha,a},-\Delta)=\frac{-2\cos(2av)+4\pi(\alpha_0+\alpha_1)a}{\pi
a v^2}+O(v^{-3}).
\]

Using the integral representation for the zeta function given in
equation (\ref{eeb}), we can split the integral at $x=1$
\begin{align*}
\zeta(s;-\Delta_{\alpha,a},-\Delta)=\zeta_0(s;a)+\zeta_\infty(s;a)
&=\int_0^1 v^{-2s} e(v;-\Delta_{\alpha,a},-\Delta)dv
+\int_1^\infty v^{-2s} e(v;-\Delta_{\alpha,a},-\Delta)dv.
\end{align*}

Making use of  the above expansion of the function
$e(v;-\Delta_{\alpha,a},-\Delta)$ for small $v$, we see that
$\zeta_0(s;a)$ is regular near $s=-\frac{1}{2}$, and its value is
\[
\zeta_0\big(-\frac{1}{2};a\big)=\int_0^1 v
e(v;-\Delta_{\alpha,a},-\Delta)dv.
\]

Next, $\zeta_\infty(s;a)$ is not regular near $s=-\frac{1}{2}$. However, using the asymptotic expansion given above
\begin{align*}
\zeta_\infty(s;a)=z_A(s;a)+z_B(s;a)=&\int_1^\infty v^{-2s} \left(e(v;-\Delta_{\alpha,a},-\Delta)+\frac{2\cos(2av)-4\pi(\alpha_0+\alpha_1)a}{\pi av^2}\right)dv\\
&-\int_1^\infty v^{-2s} \frac{2\cos(2av)-4\pi(\alpha_0+\alpha_1)a}{\pi av^2}dv.
\end{align*}

Using the expansion of the function $e(v;-\Delta_{\alpha,a},-\Delta)$ for large $v$, we see that
that $z_A(s;a)$ is regular at $s=-\frac{1}{2}$ and its value is
\[
z_A\big(-\frac{1}{2};a\big)=\int_1^\infty v
\left(e(v;-\Delta_{\alpha,a},-\Delta)+\frac{2\cos(2av)-4\pi(\alpha_0+\alpha_1)a}{\pi
av^2}\right)dv.
\]

The last term $z_B(s;a)$ is not regular at $s=-\frac{1}{2}$. However, we can deal with this term exactly:
\begin{align*}
z_B(s;a)=&-\int_1^\infty v^{-2s} \frac{2\cos(2av)-4\pi(\alpha_0+\alpha_1)a}{\pi av^2}dv\\
=&-\frac{2}{\pi a}\int_1^\infty v^{-2s}
\frac{\cos(2av)}{v^2}dv+4(\alpha_0+\alpha_1)\frac{1}{2s+1},
\end{align*}
and therefore
\begin{align*}
\Ru_{s=-\frac{1}{2}} z_B(s;a)=2(\alpha_0+\alpha_1),\hspace{30pt}
\Rz_{s=-\frac{1}{2}} z_B(s;a)=\frac{2{\rm ci}(2a)}{\pi a}.
\end{align*}


As a consequence, the partition function of our model in the range
$4\pi^2\alpha_0 \alpha_1 a^2\geq 1$ is:
\begin{align*}
\log Z=&-\frac{\beta}{2}\int_1^\infty v \left(e(v;-\Delta_{\alpha,a},-\Delta)+\frac{2\cos(2av)-4\pi(\alpha_0+\alpha_1)a}{\pi av^2}\right)dv+2(\alpha_0+\alpha_1)\left(\log 2\ell-1\right)\beta\\
&-\frac{{\rm ci}(2a)}{\pi a }\beta-\frac{\beta}{2}\int_0^1 v
e(v;-\Delta_{\alpha,a},-\Delta)dv -\int_0^\infty
\log\big(1-\e^{-v\beta}\big) e(v;-\Delta_{\alpha,a},-\Delta) dv.
\end{align*}

Note that the zeta-function regularization used  implies the
presence of the renormalization scale $\ell$ in the final expression
for the canonical partition function. In the present case, the
dependence drops out as soon as one is interested in evaluation of
physical quantities, as for example the Casimir force, defined as
the derivative of the regularized vacuum  energy with respect to
external parameter $a$.

\vskip .2in

\centerline{\bf Acknowledgment}

The authors thank G. Cognola and L. Vanzo for useful conversations.

\end{document}